\documentclass[conference]{IEEEtran}
\usepackage[T1]{fontenc}% optional T1 font encoding

\def\argmax{\mathop{\rm \arg\!\max}}

\usepackage{graphicx}
\usepackage[noadjust]{cite}
\usepackage{mcite}
\usepackage{amsfonts,helvet}
\usepackage{fancyhdr}
\usepackage{threeparttable}
\usepackage{epsf,epsfig}
\usepackage{amsthm}
\usepackage{amsmath}
\usepackage{siunitx}
\usepackage{amssymb}
\usepackage{dsfont}
\usepackage{subfigure}
\usepackage{color}
 \usepackage[noend]{algorithmic}
\usepackage{enumerate}
\usepackage{cancel}
\usepackage{bbm}
\usepackage{dsfont}
\usepackage[subnum]{cases}
\usepackage{adjustbox}
\usepackage[linesnumbered,ruled]{algorithm2e}
\usepackage{multicol}
\usepackage[english]{babel}
\usepackage{array}

\newmuskip\pFqmuskip

\def\ba{{\bf a}}
\def\bbb{{\bf b}}

\def\bd{{\bf d}}

\def\bh{{\bf h}}

\def\br{{\bf r}}
\def\bs{{\bf s}}

\def\bw{{\bf w}}
\def\bx{{\bf x}}
\def\by{{\bf y}}
\def\bz{{\bf z}}

\def\bA{{\bf A}}

\def\bH{{\bf H}}

\def\bW{{\bf W}}

\def\cC{\mbox{$\mathcal{C}$}}

\def\cI{\mbox{$\mathcal{I}$}}

\def\cK{\mbox{$\mathcal{K}$}}

\def\cN{\mbox{$\mathcal{N}$}}

\def\cQ{\mbox{$\mathcal{Q}$}}

\def\bbC{\mbox{$\mathbb{C}$}}

\def\bbE{\mbox{$\mathbb{E}$}}

\def\bbP{\mbox{$\mathbb{P}$}}
\def\bbQ{\mbox{$\mathbb{Q}$}}

\def\bb1{\mathbbm{1}}

\def\bSigma{\boldsymbol{\Sigma}}

\IEEEoverridecommandlockouts

\title{\huge
Adaptive Learning-Based Detection for \\ One-Bit Quantized Massive MIMO Systems
}

\author{Yunseong Cho, Jinseok Choi$^\dagger$, and Brian L. Evans \\
\IEEEauthorblockA{\normalsize{6G@UT Research Center, Wireless Networking and Communications Group, The University of Texas at Austin}\\
Email: yscho@utexas.edu, bevans@ece.utexas.edu}
$^\dagger$Dept. of Electrical Engineering, Ulsan National Institute of Science and Technology \\
Email: jinseokchoi@unist.ac.kr
}
\begin{document}
\maketitle

%%%%%%%%%%%%%%%%%%%%%%%%%%%%%%%%%%%%%%%%%%%%%%%%%%%%%%%%
\begin{abstract}
We propose an adaptive learning-based framework for uplink massive multiple-input multiple-output (MIMO) systems with one-bit analog-to-digital converters.
Learning-based detection does not need to estimate channels, which overcomes a key drawback in one-bit quantized systems.
During training, learning-based detection suffers at high signal-to-noise ratio (SNR) because observations will be biased to $+1$ or $-1$ which leads to many zero-valued empirical likelihood functions.
At low SNR, observations vary frequently in value but the high noise power makes capturing the effect of the channel difficult.
To address these drawbacks, we propose an adaptive dithering-and-learning method.
During training, received values are mixed with dithering noise whose statistics are known to the base station, and the dithering noise power is updated for each antenna element depending on the observed pattern of the output.
We then use the refined probabilities in the one-bit maximum likelihood detection rule.
Simulation results validate the detection performance of the proposed method vs. our previous method using fixed dithering noise power as well as zero-forcing and optimal ML detection both of which assume perfect channel knowledge.
\end{abstract}
%%%%%%%%%%%%%%%%%%%%%%%%%%%%%%%%%%%%%%%%%%%%%%%%%%%%%%%%

% \blfootnote{This fixed dithering noise power approach was presented in \cite{choi2019robust}.}

%%%%%%%%%%%%%%%%%%%%%%%%%%%%%%%%%%%%%%%%%%%%%%%%%%%%%%%
\section{Introduction}
\label{sec:intro}
%%%%%%%%%%%%%%%%%%%%%%%%%%%%%%%%%%%%%%%%%%%%%%%%%%%%%%%

% Massive MIMO one-bit
Positioning large antenna arrays has been considered as one of the emerging technologies for future communications such as massive multiple-input-multiple-output MIMO for sub-6GHz systems \cite{ngo2013energy, larsson2014massive} and millimeter wave communications \cite{pi2011introduction, andrews2014will, heath2018foundations}.
Due to the small wavelength of mmWave signals and small antenna spacing, the mmWave system allows the installation of more antennas per unit area, and hence a large number of high-precision analog-to-digital converters (ADCs) at receivers causes a prohibitively huge power consumption, which becomes the main bottleneck in the realistic deployment because a high-resolution ADC is particularly power-hungry as the power
consumption of an ADC is scaled exponentially with the number of quantization bits.
To overcome this issue, deploying low-precision ADCs has been evaluated as a low-power solution over recent years \cite{wen2016bayes,studer2016quantized,choi2017resolution,choi2018antenna,choi2021quantized}. 
The one-bit data converter is particularly attractive due to its ability to enhance power efficiency, lower hardware cost, and simplify analog processing in receivers \cite{mezghani2007ultra, mo2015capacity,wang2015multiuser, choi2015quantized, choi2016near,cho2019one,jeon2018one,park2021construction,mollen2017uplink}.

%%%%%%%%%%%%%%%%%%%%%%%%%%%%%%%%%%
\begin{figure}[!t]
    \centering
    \includegraphics[width=0.95\columnwidth]{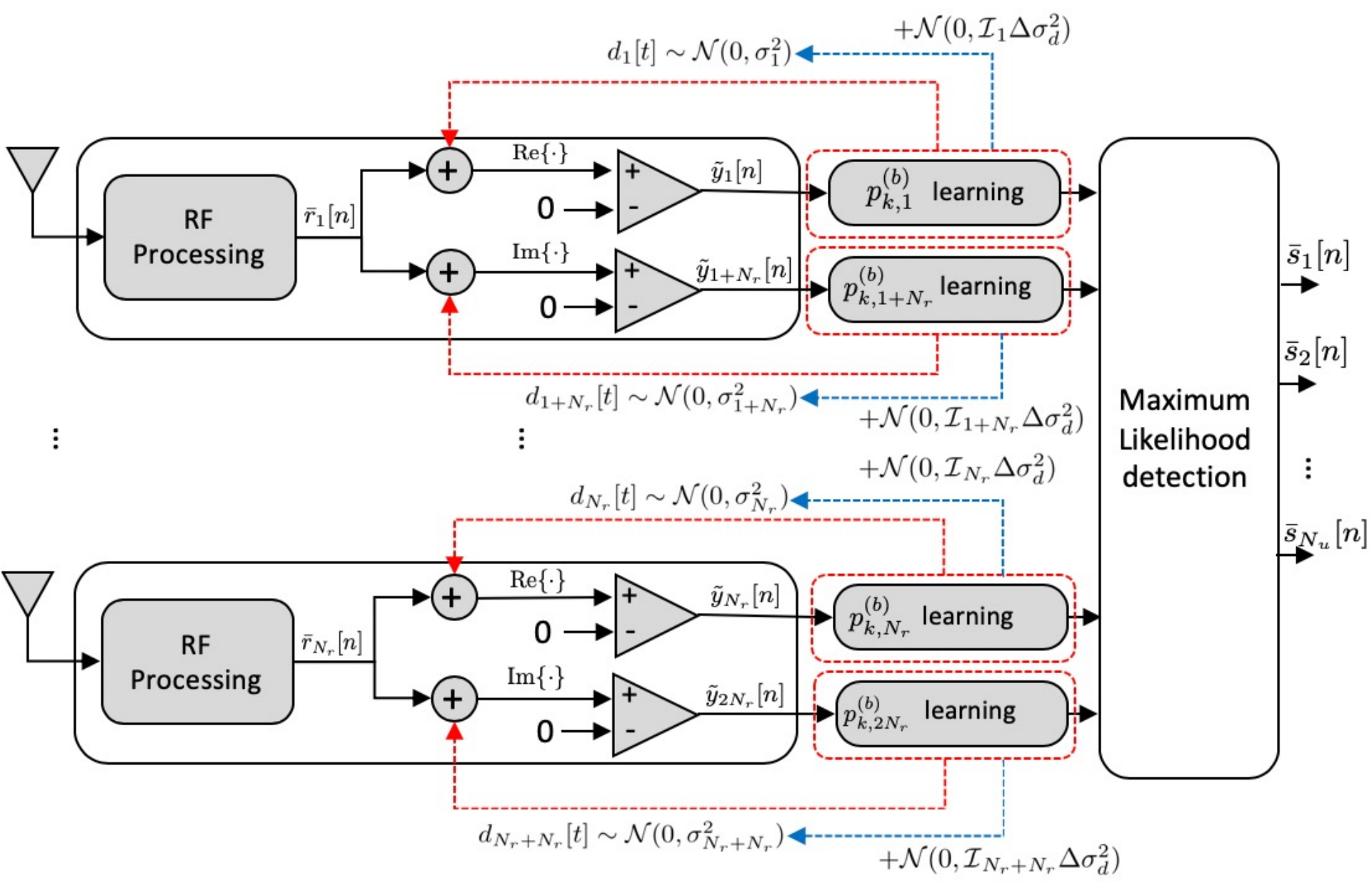}
    \vspace{-0.2cm}
    \caption{A receiver architecture for the training phase with a dithering signal (in red). The BS decides whether to add intensity to the dither power based on the observation patterns (in blue).}
    \vspace{-0.2cm}
    \label{fig:receiver}
\end{figure}
%%%%%%%%%%%%%%%%%%%%%%%%%%%%%%%%%%
% One-bit ADC papers - channel information & Intro to learning-based detection

Several modern one-bit channel estimation, detection, and beamforming techniques have been introduced \cite{wang2015multiuser, choi2015quantized, choi2016near,cho2019one,jeon2018one,park2021construction}.
Low-complexity symbol-level beamforming methods for one-bit quantized systems were developed for quadrature-amplitude-modulation (QAM) constellations \cite{park2021construction}.
Taking into account the heavily quantized signals and antenna correlation, the authors in \cite{wang2015multiuser} devised an iterative multiuser detection by using a message-passing de-quantization algorithm.
For MIMO systems, the optimal maximum likelihood (ML) detector was introduced and a near-optimal ML detector was also proposed by transforming the ML detection problem into a tractable convex optimization problem \cite{choi2016near}.
In \cite{cho2019one}, the authors presented successive-interference-cancellation one-bit receiver which can be applied to modern channel coding techniques.
However, such detection methods require perfect channel state information (CSI), which is unrealistic with one-bit quantized signals.
Various channel estimation methods were developed such as least-squares (LS), ML, zero-forcing (ZF), and Bussgang decomposition-based methods \cite{choi2016near,li2017channel}; however, channel estimation with one-bit quantized signals still suffers degradation in estimation accuracy compared to high-precision ADC systems.
In this context, we investigate a learning-based detection that replaces one-bit channel estimation with an ML probability learning process.  

Several learning-based techniques have recently been investigated \cite{nguyen2021svm,jeon2018supervised,jeon2018reinforcement}.
Support vector machines \cite{nguyen2021svm} were used for efficient channel estimation and data detection with one-bit quantized observations.
The authors in \cite{jeon2018one} applied sphere decoding to the one-bit quantized system to reduce complexity while achieving near-optimal detection performance.
Viewing the one-bit ADC systems as a classification problem, various supervised-learning-based data detection techniques were provided by estimating effective channels and learning the non-linear system response \cite{jeon2018supervised}.
However, such methods are heavily influenced by channel estimation accuracy.
Recently, 
Unlike previous learning-based approaches that focused on developing detection mechanisms based on channel estimation, we rather focus on applying one-bit ML detection and learning likelihood functions to overcome the problem of the learning process with the limited training length.

% Contributions & difference
In this work, we explore an adaptive learning-based ML detection approach that replaces one-bit channel estimation with the adaptive probability learning process in Fig.~\ref{fig:receiver}. 
We use a dithering technique to infer likelihood functions from the dithered signals, which significantly reduces the number of zero-valued likelihood functions.
To this end, we first add artificial dithering noise before quantizing the received signal, and then use de-noising to retrieve the artificial-noise-free likelihood probability. \cite{choi2019robust}
We extend our work in \cite{choi2019robust} to adapt the dithering power for each antenna element using feedback during training.
During training, we estimate the likelihood probability with an acceptable training length as the extra randomness is likely to trigger a change in sign in the sequence of the quantized signals.
Since the dithering-based algorithm is affected by the dithering power, we propose a heuristic algorithm to adjust the dithering noise power depending on the pattern of the one-bit quantized vectors.
% Accordingly, the base station (BS) can appropriately control the dithering noise.
Simulation results demonstrate that, in contrast to the conventional learning-based one-bit ML detection, the proposed adaptive learning-based detection technique exhibits more reliable detection performance and achieves comparable performance to the optimal one-bit ML detection that unrealistically requires perfect CSI.

{\it Notation}: $\bf{A}$ is a matrix and $\bf{a}$ is a column vector. 
$\mathbf{A}^T$  denotes transpose operation. 
We denote $a_{i}$ as the $i$th element of $\bf a$. 
With mean $\mu$ and variance $\sigma^2$, we generate a real Gaussian distribution and a complex Gaussian distribution using $\mathcal{N}(\mu, \sigma^2)$ and $\mathcal{CN}(\mu, \sigma^2)$, respectively. 
$\rm diag (\bf a)$ creates a diagonal matrix that has $a_i$'s as its diagonal entries. 
${\bf I}_N$ denotes a $N\times N$ identity matrix.
${\bf 1}_N$ and ${\bf 0}_N$ are a $N \times 1$ one vector and zero vector, respectively.
${\rm Re}\{\bA\}$ and ${\rm Im}\{\bA\}$ take the real and imaginary part of $\bA$, respectively.
$\bb1{\{A\}}$ is the indicator function which outputs 1 if $A$ is true, and 0 otherwise.
$\bbE[\cdot]$ is the expectation operator.
For a complex-based vector and matrix, we define real-valued expansions as $\omega(\ba)= \begin{bmatrix} {\rm Re}\{\ba\} \\  {\rm Im}\{\ba\}\end{bmatrix}$ and $\omega(\bA)=\begin{bmatrix} {\rm Re}\{\bA\} & -{\rm Im}\{\bA\} \\ {\rm Im}\{\bA\} & {\rm Re}\{\bA\}\end{bmatrix}$, respectively.

%%%%%%%%%%%%%%%%%%%%%%%%%%%%%%%%%%%%%%%%%%%%%%%%%%%%%%%
\section{Preliminaries}
\label{sec:system}
%%%%%%%%%%%%%%%%%%%%%%%%%%%%%%%%%%%%%%%%%%%%%%%%%%%%%%%

\subsection{System Model}

We consider uplink multiuser MIMO communication systems where the BS has $N_r$ receive antennas and concurrently communicates with $N_u$ single-antenna user devices.
For our massive MIMO systems, we assume $N_r \gg N_u$.
A block fading channel is assumed where a channel matrix is invariant during $N_c$ time slots.
We then split the uplink transmission into training with $N_t$ time slots and data transmission with $N_d$ slots, i.e., $N_c=N_t+N_d$.
During training, users transmit pilot symbols up to $N_{t}$ symbol times in total.
We use $K$ to denote the number of possible symbol combinations that users can send together and $N_{tr}$ to represent the number of transmissions of each signal, which implies $N_t \geq KN_{tr}$.

Let $\bar \bs[t] \in \bbC^{N_u}$, $t= 1,\dots, N_c$, denote a symbol vector at time $t$.
Then, the received signal vector at time $t$ is
\begin{align}
    \label{eq:r}
    \bar\br[t] = \sqrt{\rho}\bar\bH\bar\bs[t] + \bar\bz[t],
\end{align}
where $\bar\bH \in \bbC^{N_r\times N_u}$ is the channel matrix between the BS and $N_u$ users, whose row vector $\bar\bh_i^T$ indicates the channel vector between all users and the $i$th BS antenna element.
Let $\bbQ_M$ denote the set of $M$-ary QAM constellation points from which $\bar s_u[t]$ is generated where $\bar s_u[t]$ denotes the $u$th element of $\bar \bs[t]$ and symbol of the $u$th user.
$\bar s_u[t]\in\bbQ_M$ has zero mean and unit variance, i.e.,  $\bbE[\bar s_u] =0$ and $\bbE[|\bar s_u[t]|^2] = 1$.
$\bar\bz[t]$ is the additive noise vector at time $n$ that follows $\mathcal{CN}\left({\bf 0}_{N_r}, {N_0}{\bf I}_{N_r}\right)$ with noise variance $N_0$.
We define SNR as $\gamma = \rho/N_0$.

The real and imaginary components of the received signal are each quantized with one-bit ADCs to capture the sign, i.e., $+1$ or $-1$.
The quantized signal can be represented as
\begin{align}
    \bar\by[t] = \cQ({\rm Re}\{\bar \br[t]\}) + j\cQ({\rm Im}\{\bar \br[t]\})
\end{align}
where $\cQ(\cdot)$ is a element-wise one-bit quantizer.
The received signal in the complex-vector form $\bar \br$ can be rewritten in a real-valued vector representation as
\begin{align}
    \br[t] = \omega(\bar\br[t]) = \sqrt{\rho}\bH\bs[t] + \bz[t]
\end{align}
where $\bH=\omega(\bar\bH)$, $\bs[t]=\omega(\bar\bs[t])$, and $\bz[t]=\omega(\bar\bz[t])$.
% \begin{gather}
%     \bH = \begin{bmatrix} {\rm Re}\{\bar\bH\} & -{\rm Im}\{\bar\bH\} \\ {\rm Im}\{\bar\bH\} & {\rm Re}\{\bar\bH\}\end{bmatrix},    \\
%     \bs[t] = \begin{bmatrix} {\rm Re}\{\bar\bs[t]\} \\  {\rm Im}\{\bar\bs[t]\}\end{bmatrix}, \  \bz[t] = \begin{bmatrix} {\rm Re}\{\bar\bz[t]\} \\  {\rm Im}\{\bar\bz[t]\}\end{bmatrix}. 
% \end{gather}
We rewrite the quantized signal in a real-vector form as
\begin{align}
    \by[t] & = \cQ(\br[t]) = \cQ(\sqrt{\rho}\bH\bs[t] + \bz[t]),
\end{align}
and each value $r_i[t]$ is quantized to be $y_i[t] = +1$ if $r_i[t] \geq 0$ or $y_i[t] = -1$ otherwise. 

\subsection{One-Bit ML Detection with CSI}
\label{subsec:optML}

We first introduce the conventional one-bit ML detection with the full CSI.
Since each user sends one of $M$ possible signals, the number of possible symbol vectors from $N_u$ users is $K=M^{N_u}$. 
We define the index set of all possible symbol vectors as $\cK=\{1,\ldots,K\}$ and use $\bs_k$ to denote the $k$th pilot symbol vector in a real-vector form.
Let $p_{k}(\beta)$ where $\beta\in\{-1,+1\}$ denote the probability that $i$th antenna component receives $\beta$ when users transmit $k$th symbol vector.
Assuming uncorrelated antennas, the likelihood probability of the quantized signal vector $\by[t]$ for a given channel $\bH$ and transmit symbol vector $\bs_k$ is given as
\begin{align}
	\label{eq:ML}
    \bbP(\by[t]|\bH, \bs_k) = \prod_{i=1}^{2N_r}p_{k}(y_i[t])
\end{align}
where the likelihood function for $i$th antenna element for an observation $y_i[t] \in \{-1,+1\}$ with the full CSI is defined as
\begin{align}
    \label{eq:p}
   p_k(y_i[t])\! &=\!  \bbP(y_i[t]|{\bh}_i, \bs_k\!) \\ 
   &= \Phi\left(y_i[t]  \psi_{k,i}\right),
\end{align}
where $\psi_{k,i}=\sqrt{\frac{\rho}{N_0/2}}\bh^T_i\bs_k$ is the effective channel when transmitting $k$th symbol vector and $\Phi(x) = \int_{-\infty}^{x} \frac{1}{\sqrt{2\pi}}e^{-{\tau^2}/{2}}d\tau$ is the cumulative distribution function of a standard Gaussian distribution.
Based on \eqref{eq:ML}, the one-bit ML detection rule is given as
\begin{align}
    \label{eq:MLD}
    k^\star[t] = \argmax_{k \in \cK} \prod_{i=1}^{2N_r}p_{k}(y_i[t])
\end{align}
The detected symbol vector is defined as $\hat\bs[t]$ = $\bs_{k^\star[t]}$.
However, the detection rule in \eqref{eq:MLD} requires full CSI to compute \eqref{eq:p}, which is intractable when employing one-bit ADCs.

\subsection{One-bit ML Detection without CSI (Naive Approach)}

\label{subsec:ML_learning}
%%%%%%%%%%%%%%%%%%%%%%%%%%%%%%%%%%%%%%%%%%%%%%%%%%%%%%%

Now, we outline a straightforward learning strategy that does not require channel estimation, however, does require $N_{tr}$ training sequences.
Each pilot symbol vector $\bs_k\in\bbQ_M^{Nu}$ is transmitted $N_{tr}$ times throughout the pilot transmission of length $N_t$. 
The BS learns likelihood functions by observing the frequency of $y_i[t] = 1$ and $y_i[t] = -1$ as 
\begin{align}
    \label{eq:p1_learning}
    \hat p^{(\beta)}_{k,i}\! =\! \begin{cases}  \hat p^{(+1)}_{k,i} = \frac{1}{N_{tr}}\sum_{t=1}^{N_{tr}} \bb1{\{y_i[(k-1)N_{tr} + t] = 1\}}\\ 
     \hat p^{(-1)}_{k,i} =  1- \hat p^{(+1)}_{k,i} 
    \end{cases}
\end{align}
where $\beta \in \{+1,-1\}$.
The operation in \eqref{eq:p1_learning} measures the number of $+1$'s at $i$th antenna element out of the $N_{tr}$ observations triggered by $\bs_k$.
After learning the likelihood functions, the BS obtains the estimate of the likelihood probability for a given data signal $\by[t]$ as 
\begin{align}
    \label{eq:Py_learning} 
    \bbP(\by[t]|\bH,\bs_k) \!\approx\! \prod_{i=1}^{2N_r}\!\!\Big(\hat p^{(+1)}_{k,i}\bb1\{y_i[t] \!=\! 1\} \!+ \!\hat p^{(-1)}_{k,i}\bb1\{y_i[t] \!=\! -1\}\!\Big),
\end{align}
and the receiver can perform the ML detection in \eqref{eq:MLD} by searching the best index that maximizes \eqref{eq:Py_learning} over the $K$ possible symbol vectors.  

Although such a one-bit ML approach can provide a near-optimal detection performance with the simple function learning, it may suffer from critical performance degradation with a limited length of training;
in the high SNR, the $N_{tr}$ observations of each antenna repeatedly observe either $+1$ or $-1$ due to low variance aggregate noise.
This phenomenon further leads to a number of zero empirical likelihood functions in \eqref{eq:p1_learning}, e.g., $\hat p^{(\beta)}_{k,i} = 0$.
This is because the one-bit quantized observation in the high SNR becomes so deterministic that it is hard to observe a change in the sign of the quantized output sequences during the $N_{tr}$ transmissions of the symbol vector $\bs_k$. 
We name it an undertrained likelihood function, which completely ruins the ML detection rule since any zero probability can cancel out the whole product in \eqref{eq:ML}.
To overcome such a limitation, we propose a novel learning-based one-bit ML detection method that does not require explicit CSI and is robust to the length of the training sequences in the following section.

%%%%%%%%%%%%%%%%%%%%%%%%%%%%%%%%%%%%%%%%%%%%%%%%%%%%%%%
\section{Adaptive Statistical Learning without CSI}
\label{sec:ML}
%%%%%%%%%%%%%%%%%%%%%%%%%%%%%%%%%%%%%%%%%%%%%%%%%%%%%%%

We present an adaptive learning-based ML detection method for one-bit ADC systems in order to achieve optimal ML detection performance without requiring explicit estimation of wireless channels.
% Being identical to the maximum a posteriori estimation, the ML estimation is optimal in minimizing the probability of detection error when all possible transmit symbols have an equal probability of being transmitted.

%%%%%%%%%%%%%%%%%%%%%%%%%%%%%%%%%%%%%%%%%%%%%%%%%%%%%%%

%%%%%%%%%%%%%%%%%%%%%%%%%%%%%%%%%%%%%%%%%%%%%%%%%%%%%%%

\subsection{Incremental Dither-and-Learning with $N$-steps}
\label{sec:DL}

To resolve the problem caused by the undertrained likelihood functions, we propose the incremental dither-and-learning (iDL) method that can learn the likelihood functions with a reasonable $N_{tr}$ without CSI. 
As shown in Fig.~\ref{fig:receiver}, the BS appends dithering signals $d_i[t]$ to $r_i[t]$ during the training phase.
After placing the additional noise, the quantization input in the real-vector form becomes
\begin{align}
    \br_{d,k}[t] &= \br_k[t] + \bd[t]\\
    & = \sqrt{\rho}\bH\bs_k + \bz[t] + \bd[t].
\end{align}
We use $\sigma_{d,i}^2$ to denote the dithering power at $i$th antenna and assume $\bd[t] \sim \mathcal{N}({\bf 0}_{2N_r}, \bSigma)$ where $\bSigma = {\rm diag}(\sigma_{d,1}^2,\ldots,\sigma_{d,2N_r}^2)$ represents the collection of dithering powers.
Note that a small $\sigma_{d,i}^2$ still triggers undertrained likelihood functions while a larger $\sigma_{d,i}^2$ hinders extracting the symbol information as noise term becomes dominant.
Then, the dithered and quantized signal associated with $k$ symbol vector becomes
\begin{align}
    \by_{d,k}[t] = \cQ(\sqrt{\rho}\bH\bs_k + \bz[t] + \bd[t])\in\{+1,-1\}^{2N_r}.
\end{align}

The BS computes the estimated likelihood function for the dithered signals $\hat p_{k,i}^{(\beta)}$ as in \eqref{eq:p1_learning} for $\beta \in \{+1, -1\}$.
Without loss of generality, let us fix $\beta = +1$ for simplicity. Then, as shown in \eqref{eq:p}, $\hat p_{k,i}^{(+1)}$ is  theoretically derived as
\begin{align}
    \label{eq:p_dither}
    \hat p_{k,i}^{(+1)} \approx \Phi\left(\sqrt{\frac{2\rho}{N_0 + \sigma^2 }}\bh_i^T\bs_k\right),
\end{align}
Since $N_0$ and $\sigma_d^2$ are known to the BS and $\tilde p^{(+1)}_{k,i}$ is learned from \eqref{eq:p1_learning}, the BS can find the estimate of $\psi_{k,i} \triangleq \sqrt{\frac{\rho}{N_0/2}}\bh_i^T\bs_k$ by using \eqref{eq:p_dither} and the de-nosing phase defined as
\begin{align}
	\label{eq:effective_channel}
    \hat{\psi}_{k,i} &= \sqrt{\frac{\rho}{N_0/2}}\tilde{\bh}_i^T\bs_k  \\
    &= \sqrt{1+ \frac{\sigma^2}{N_0}}\Phi^{-1}\left(\hat p_{k,i}^{(+1)}\right).
\end{align}

Finally, the BS exploits the estimated effective channel $\hat{\psi}_{k,i}$ and known or estimated $N_0$ to approximate the true (non-dithered) likelihood function $p_{k,i}^{(+1)}$ by using \eqref{eq:p} as $\Phi\left(  \hat{\psi}_{k,i}\right)$.
Since the likelihood function of the dithered signal $\hat p_{k,i}^{(+1)}$ in \eqref{eq:p_dither} is much less likely to have zero probability compared with the non-dithered case, the BS can learn the majority of the likelihood functions $\hat p^{(+1)}_{k,i}$ with a reasonable training length.

The undertrained likelihood functions are undesirable because we lead to $\bbP(\by[t]|\bH,\bs_k) = 0$ for many candidate symbols $\bs_k$, which may exclude the desired symbol from the ML detection. However, the fixed dithering variance does not suitably adjust the dithering power and this behavior can cause two fundamental problems:
1) When the dithering power is low and the SNR is equivalently high, there still exists lots of undertrained likelihood functions in spite of the increased variance, however, the BS cannot do further; and 
2) In addition, with a high dithering power, even if the training framework rarely shows zero probabilities, it is demanding to capture the effect of the channel because the noise power becomes dominant. 
Therefore, the BS has to support an acceptable dithering power based on the behavior of received observations.
To prevent the problems, we propose an incremental dither-and-learning (iDL) method that fits the dithering power into a proper range.

\begin{figure}[!t]
    \centering
    \includegraphics[width= 1  \columnwidth]{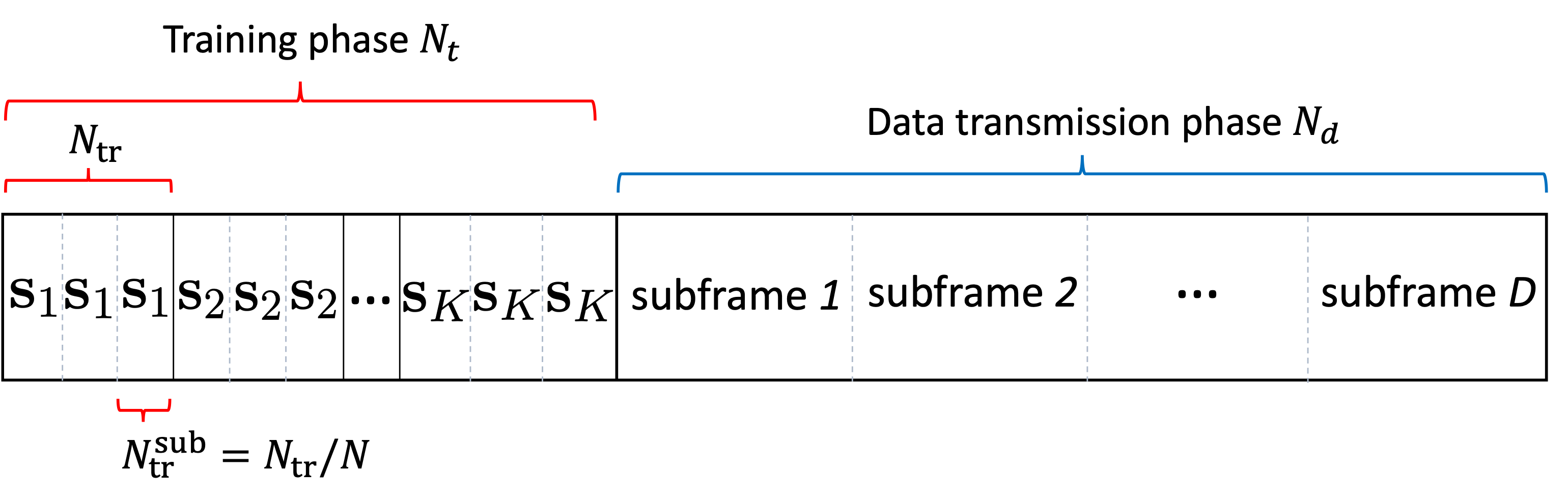}
    \caption{Illustration of communication frame composed of a training phase with $N$ sub-blocks and a data transmission phase.}
    \label{fig:frame}
\end{figure}

\begin{figure}[!t]
    \centering
    \includegraphics[width= 1\columnwidth]{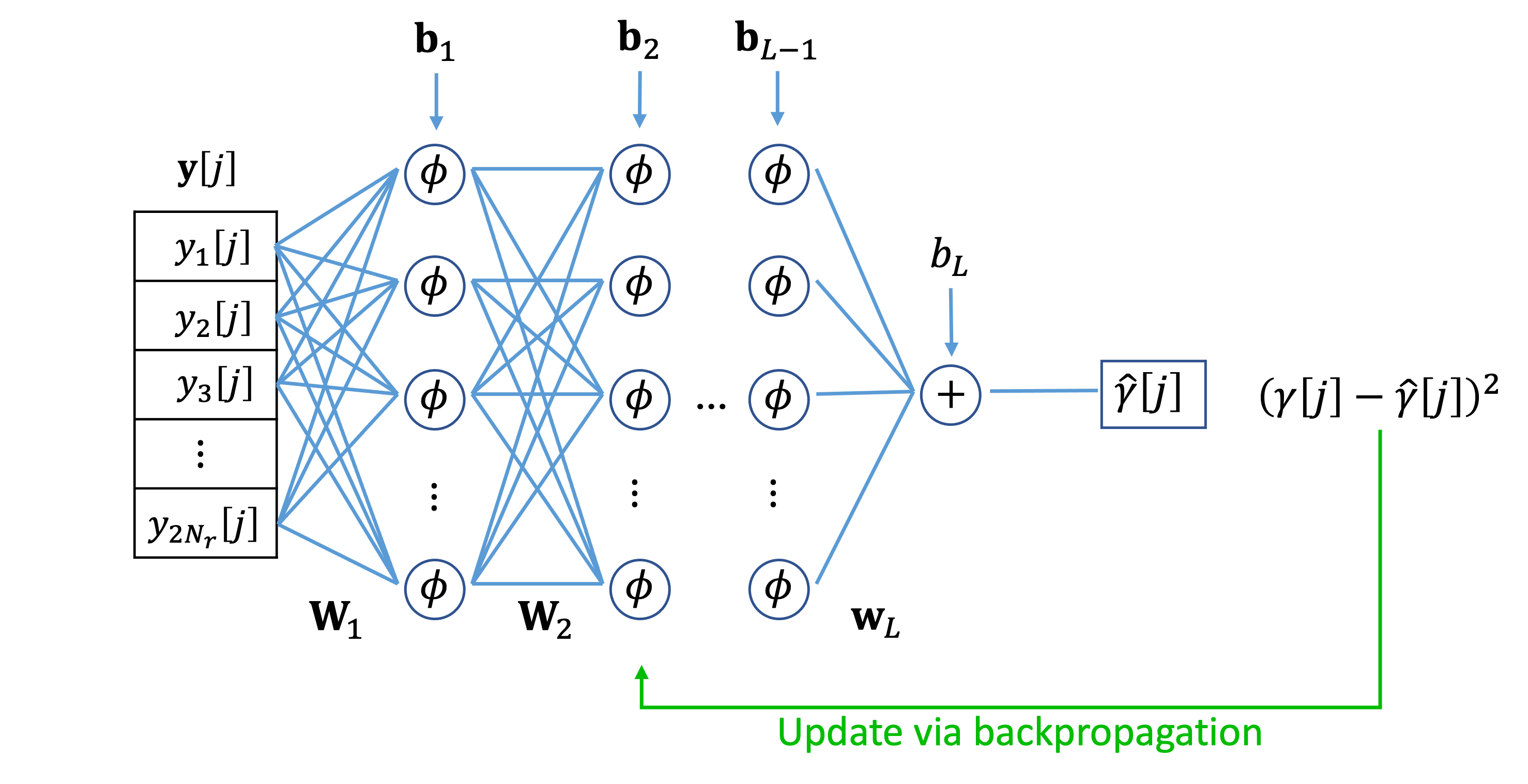}
    \vspace{-0.3cm}
    \caption{Illustration of the SNR offline training via deep neural network.}
    \label{fig:DNN}
    \vspace{-0.5cm}
\end{figure}

As shown in Fig.~\ref{fig:frame}, we first divide the $N_{tr}$ observations of $\bs_k$ into $N$ disjoint sub-blocks in which each sub-block accommodates $N_{tr}^{sub}=N_{tr}/N$ training samples where $N_{tr}$ has to be a multiple of $N$. Then, the $n$th dithered and quantized sub-block observed at $i$th antenna when transmitting $\bs_k$ can be represented as
\begin{align}
&\tilde\by_{d,k,i,n} \nonumber \\ 
&=\{ y_{d,k,i}\big[1+(k-1)N_{tr}+\left(n-1\right)N_{tr}^{sub}\big], \nonumber \\ 
&\;\ldots, y_{d,k,i}\big[(k-1)N_{tr}+nN_{tr}^{sub}\big] \}^T\in\{+1,-1\}^{N_{tr}^{sub}}, \label{eq:subblock}
\end{align}
where $n\in\{1,\ldots,N\}$ and $y_{d,k,i}[t]$ denotes the dithered observation at the $i$th antenna at time $t$ when users send $k$th symbol vector. The ML detection is not able to have an attractive performance if any of $p_{k,i}^{(+1)}$ and $p_{k,i}^{(-1)}$ is trained as zero, which means the received training sequence has the pattern of all zeros or all ones. 
% Each $n$th sub-block computes the likelihood probability using the formula introduced in Section \ref{sec:DL}.
%In this case, the training phase reports zero likelihood functions since it is not able to capture any sign changes. 

To prohibit the same problem at $(n+1)$st sub-block, we intentionally increase the dithering noise variance of $i$th antenna by $\Delta\sigma_d^2$ if $n$th sub-block outputs either $+{\bf 1}_{N_{tr}^{sub}}$ or $-{\bf 1}_{N_{tr}^{sub}}$, so that the subsequent training sequence is more likely to observe any sign changes than the previous sequence. 
However, if $n$th sub-block of $i$th antenna already reports any sign transition due to the current dithering power, it indicates that the current SNR is sufficient not to observe zero probabilities, the BS hence no longer updates the dithering noise. 
The described work can be addressed by introducing the indicator function $\cI_i$ defined for $i$th antenna and the corresponding update operation, i.e., $\cI_i\Delta\sigma_d^2$.
The indicator $\cI_i$ is initialized and maintained as 1 and the indicator is finally set to be zero to interrupt the further update of the dithering level if the sub-block starts to observe the sign changes.
Even though $\cI_i$ is set to zero, the indicator can revert to one if sign flips are not observed again.

During each $n$th sub-block, we add intensity to the dithering noise by $\cI_i\Delta\sigma_d^2$.
Therefore, the BS can maintain the proper SNR region of $i$th antenna by using the update and stopping criterion based on received observations.
Upon completing all sub-blocks, the likelihood probability of $k$-th symbol vector is determined as the mean of the likelihood probabilities among all $N$ sub-blocks associated with symbol vector $k$.
The whole process is summarized in Algorithm 1.

%%%%%%%%%%%%%%%%%%%%%%%%%%%%%%%%%%%%%%%%%%%%%%%%%%
\begin{algorithm}[t]
 \caption{Incremental Dither-and-Learning (iDL)}
 \begin{algorithmic}[1]
% \renewcommand{\algorithmicrequire}{\textbf{Input:}}
% \renewcommand{\algorithmicensure}{\textbf{Output:}}
% \REQUIRE in
% \ENSURE  out
 %\\ \textit{Initialisation} :
\STATE Initialize $p_{k,i}^{(+1)}=0 \;\;\forall k,i$ 
\STATE Fix the increase of the dithering variance, 
$\Delta\sigma_d^2$.
  \FOR {$k = 1$ to $K$}
  \STATE Initialize dithering variance as $\sigma_{d,i}^2 =0$.
  \FOR {$n=1$ to $N$}
   \FOR {$i=1$ to $2N_r$}
  \STATE Observe $\tilde\by_{d,k,i,n}$ for $n$th sub-block in \eqref{eq:subblock}
  \STATE Compute the $\hat p^{(+1)}_{k,i}$ of $\tilde\by_{d,k,i,n}$ using \eqref{eq:p1_learning}
  \STATE Derive $\hat p_{k,i}^{(+1)}$ in \eqref{eq:p_dither} and compute $\hat{\psi}_{k,i}$ in \eqref{eq:effective_channel}
  \STATE $p_{k,i}^{(+1)} \leftarrow p_{k,i}^{(+1)}+\Phi\left(\hat{\psi}_{k,i}\right)/N$
  \IF {$\tilde\by_{d,i}$ is either $+{\bf 1}_{N_{tr}^{sub}}$ or $-{\bf 1}_{N_{tr}^{sub}}$}
  \STATE $\cI_i\leftarrow 1$
  \ELSE 
  \STATE $\cI_i\leftarrow 0$ %$\;\;\;\;\;\;\;\;\;\;\;\;\;\;\rhd$ No more update of $\sigma_{d,i}^2$
  \ENDIF
  \STATE $\sigma_{d,i}^2 \leftarrow \sigma_{d,i}^2 + \cI_i\Delta\sigma_d^2$
  
  \ENDFOR
  \ENDFOR
  \ENDFOR
 \RETURN $p_{k,i}^{(+1)}$ and $p_{k,i}^{(-1)}=1-p_{k,i}^{(+1)} \;\;\forall k,i$
 \end{algorithmic}
 \end{algorithm}
%%%%%%%%%%%%%%%%%%%%%%%%%%%%%%%%%%%%%%%%%%%%%%%%%%

As a result, the effective SNR is decreased until we meet a proper sequence to capture the sign changes.
The wild fluctuations in the output values can be also prevented since the BS has control over the variance and the update is supervised by the BS to fit into the appropriate SNR region. 
Even though the BS keeps changing the dithering variance, we are still able to compute the refined likelihood function using \eqref{eq:p_dither} and \eqref{eq:effective_channel} since the updated variance is also perfectly known to the BS. %However, the framework entails $2N_{tr}$ additional buffers to store the intermediate value of $\sigma_{d,i}^2$ for each antenna. However, the performance of the algorithm may depend on the selection of different parameters such as the slicing size $N$, and the variance increment $\Delta\sigma_d^2$.

\subsection{SNR Estimation}
Despite of the properly controlled dithering power, the computation of likelihood functions using \ref{eq:effective_channel} requires the estimated SNR $\gamma$ or noise variance $N_0$ equivalently.
In this work, we perform the SNR estimation task by offline supervised learning using the deep neural network as shown in 
Fig.~\ref{fig:DNN}. 
The offline training first collects training data points $\{\by[j]; \gamma[j]\}$ where $\by[j]\in\{+1,-1\}^{2N_r}$ is the $j$th one-bit quantized observation and $\gamma[j]$ is the true SNR at time $j$.
Upon collecting enough samples, the BS selects a few training samples and performs the supervised offline learning that considers $\by[j]$'s as input and $\gamma[j]$'s as output to estimate.
Assuming that there exist $L$ layers, the estimated SNR is represented as the scalar output of the neural network expressed as 
\begin{equation}
    \hat{\gamma}[j] = \bw_L^T \bx_{L-1}+b_L,
    % \cdots \left(\bW_2\phi\left(\bW_1\by[j] + \bbb_1\right)+\bbb_2\right) \cdots
\end{equation}
where each intermediate vector is defined as $\bx_{\ell} = \phi\left(\bW_{\ell}\bx_{\ell-1} + \bbb_\ell\right)$ for $\ell\in\{1,\ldots,L-1\}$ with the initial point $\bx_0=\by[j]$ when $\phi(\cdot)$ is the element-wise activation function such as rectified linear unit or sigmoid function.
The deep neural network is updated by minimizing the estimation error, e.g., $(\gamma[j]-\hat{\gamma}[j])^2$, and hence estimates the SNR by extracting meaningful information of the one-bit observation such as statistical pattern and the number of zeros.

\section{Simulation Results}
\label{sec:simul}
%%%%%%%%%%%%%%%%%%%%%%%%%%%%%%%%%%%%%%%%%%%%%%%%%%%%%%%

We evaluate the performance of the proposed learning-based method in terms of the number of undertrained probabilities and symbol error probability (SER). 
We consider $N_r = 32$, $N_u = 4$ with $4$-QAM modulation, Rayleigh channels $\bar\bH$ in which each element follows $\cC\cN(0,1)$. We fix the dithering variance as $\sigma_d^2 = \rho/2$ and increment as $\Delta\sigma_d^2 = \rho/2$.

\subsection{Undertrained Likelihood Functions}

\begin{figure}[!t]
    \centering
    \includegraphics[width= 0.85\columnwidth]{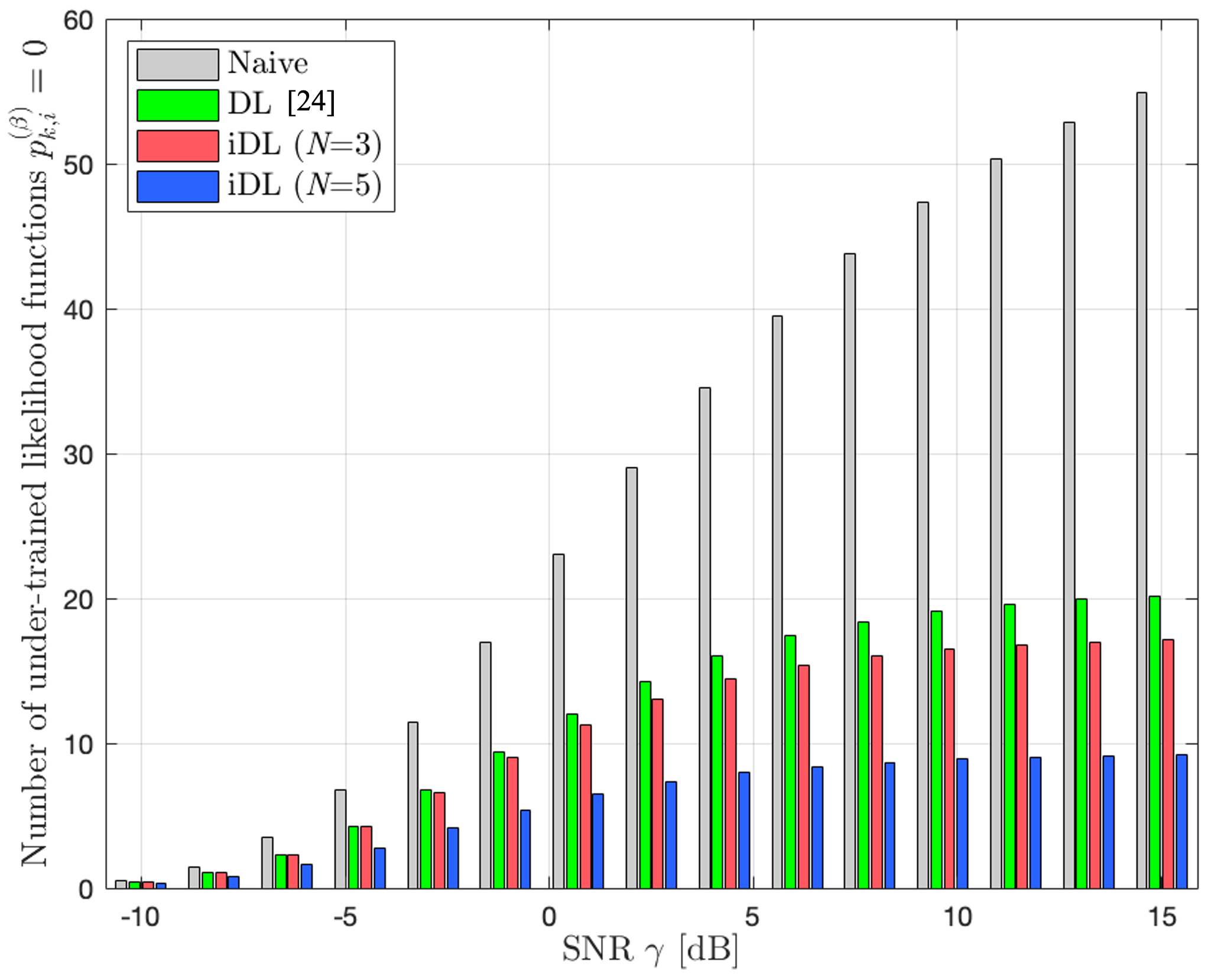}
    \caption{The number of undertrained likelihood functions out of the $2N_r$ receivers for $N_u = 4$ users, 4-QAM, $N_r = 32$ antennas, and $N_{tr}=30$ with Rayleigh channels ($\sigma_d^2 = \rho/2$ for DL and $\Delta\sigma_d^2 = \rho/2$ for iDL).}
    \label{fig:numzero}
    \vspace{-0.3cm}
\end{figure}

Fig.~\ref{fig:numzero} shows the average number of undertrained likelihood functions, i.e., $\hat p_{k,i}^{(b)} = 0$, over the wide range of the SNR levels for the non-dithering, dithering, and adaptive dithering cases with $N_{tr} = 30$.
We note that the DL method is a special case of the iDL method when all antennas exploit identical and fixed dithering power, i.e., $N=1$.
As the SNR increases, the number of undertrained likelihood functions for the non-dithering case approaches $2N_r$.
For the dithering case, however, the number of undertrained likelihood functions slowly increases with the SNR and converges to about $20$ due to the dithering effect.
Furthermore, for the incremental dithering case, the number of undertrained likelihood functions approaches $17$ and $9$ when $N$ is 3 and 5, respectively. 
Since the iDL method decides whether to increase the dithering noise depending on the realization of each sub-block, we can further optimize the learning procedure in terms of the number of undertrained likelihood functions. 
If we properly increase $N$, each antenna is more likely to avoid zero-valued likelihood probabilities. 
As a result, with dithering and slicing, the proposed algorithm can estimate much more non-zero likelihood functions, thereby increasing the detection accuracy.

%%%%%%%%%%%%%%%%%%%%%%%%%%%%%%%%%%
%%%%%%%%%%%%%%%%%%%%%%%%%%%%%%%%%%

\begin{figure}[!t]
    \centering
    \includegraphics[width=0.95\columnwidth]{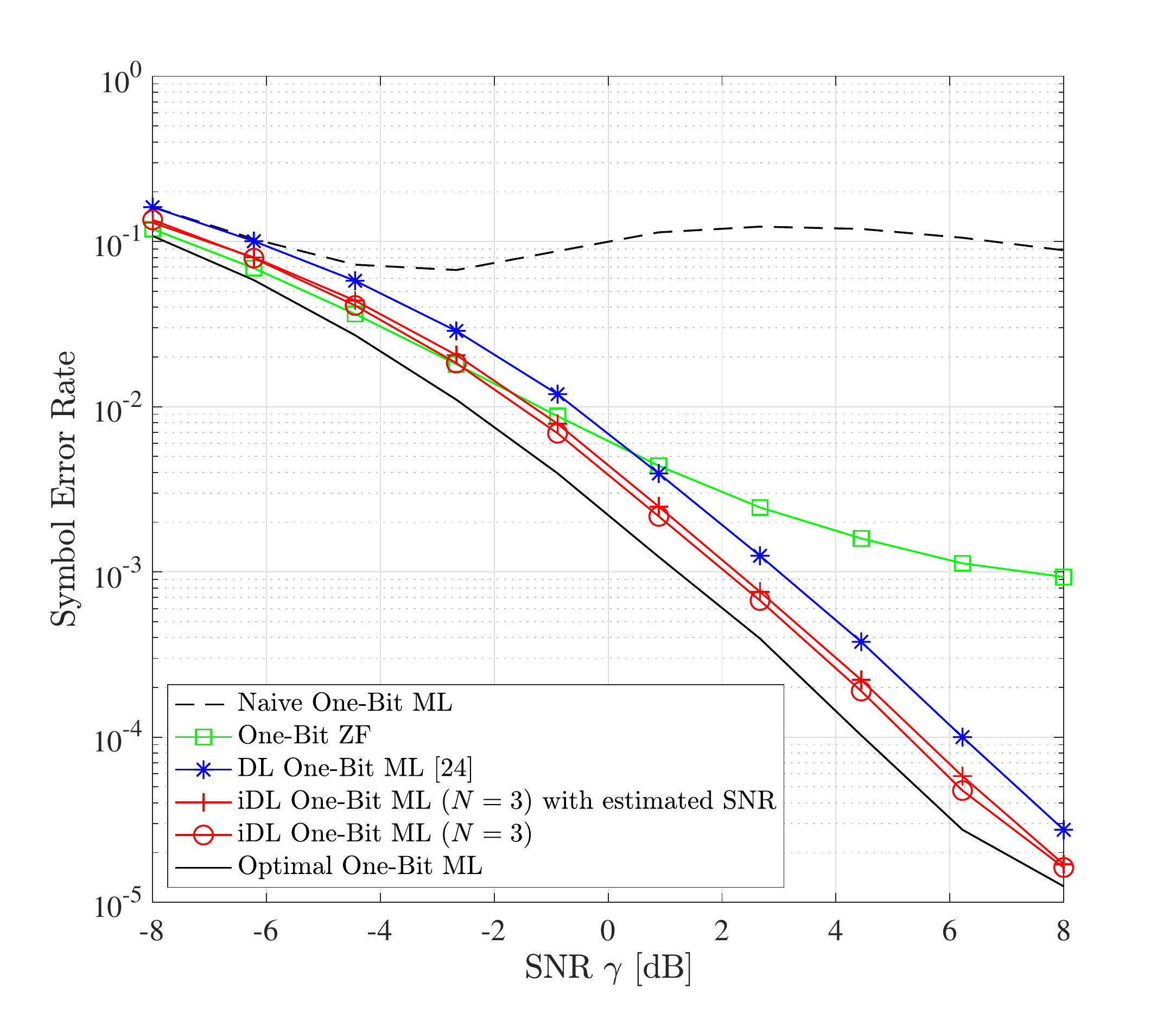}
    \vspace{-0.5cm}
    \caption{Simulation results $N_u = 4$, 4-QAM, $N_r =32$, and $N_{tr}=30$ with Rayleigh channels.}
    \vspace{-0.3cm}
    \label{fig:SER30}
\end{figure}
%%%%%%%%%%%%%%%%%%%%%%%%%%%%%%%%%%

\begin{figure}[!t]
    \centering
    \includegraphics[width=0.95\columnwidth]{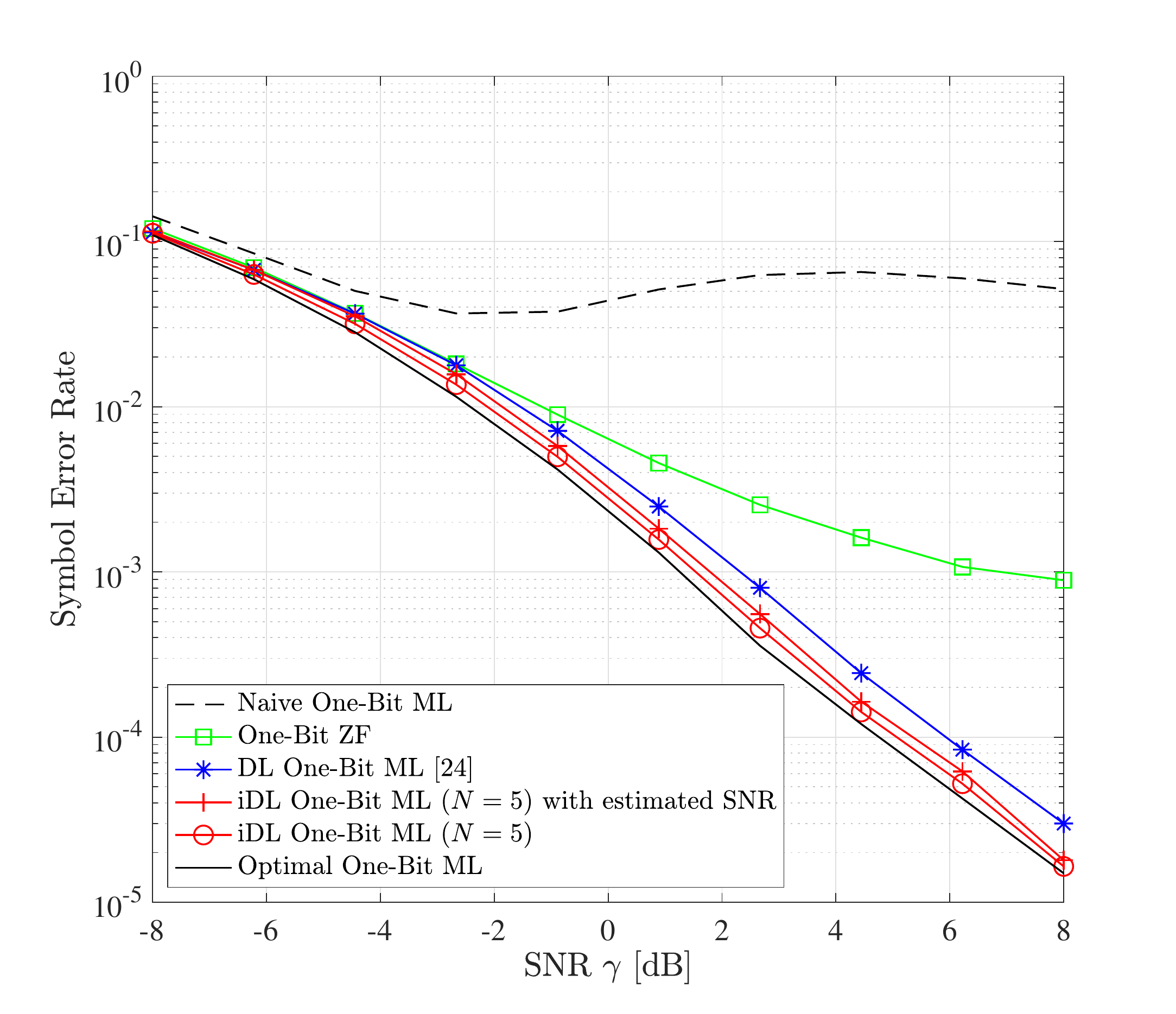}
    \vspace{-0.5cm}
    \caption{Simulation results $N_u = 4$, 4-QAM, $N_r =32$, and $N_{tr}=50$ with Rayleigh channels.}
    \vspace{-0.3cm}
    \label{fig:SER50}
\end{figure}

\subsection{Data Detection}

In the massive multiuser MIMO system, we compare the following one-bit detection methods:
\begin{enumerate}
    \item Learning one-bit ML: naive learning-based ML
    \item Dithered learning (DL) one-bit ML \cite{choi2019robust}
    \item Incremental dithered learning (iDL) one-bit ML (proposed)
    \item Incremental dithered learning (iDL) one-bit ML with estimated SNR (proposed)
    \item One-bit zero-forcing (ZF) \cite{choi2015quantized}
    \item Optimal one-bit ML
\end{enumerate}
The first four methods do not require channel estimation, however, the last two methods assume that perfect CSI is known to the BS.

Fig.~\ref{fig:SER30} illustrates the SER curves for $N_{tr} = 30$.
The one-bit ZF detection shows the large performance degradation in the medium to high SNR, and the proposed method outperforms the one-bit ZF detection.
We note that the proposed dithered case closely follows the SER performance of the optimal one-bit ML case. 
The performance improvement is achieved because the proposed method provides the data-driven adaptive likelihood function learning with the same $N_{tr}$.
In addition, the proposed iDL method has around 1.0 dB gain over the DL method by splitting the training sequence into three sub-blocks, thereby getting closer to the optimal case. 
Even though the optimal ML is implausible without explicit channel estimation, the proposed method can achieve an impressive capability without the fundamental component.
We can also notice from both simulations that the performance gap between the iDL method and the iDL with estimated SNR is marginal.
It validates the fact that the offline supervised learning can successfully capture the observation pattern to estimate the SNR required for the de-noising phase.

Fig.~\ref{fig:SER50} illustrates the SER curves for $N_{tr} = 50$.
Compared to Fig.~\ref{fig:SER30}, we can observe that the learning-based algorithms are improved and the iDL methods further approach the performance of optimal one-bit ML decoding.
The naive-learning-based one-bit ML bounces up at the higher SNR than Fig.~\ref{fig:SER30}, however, the reverse trend in the high SNR still exists because $N_{tr}=50$ is not enough to avoid undertrained likelihood probabilities.

\section{Conclusion}
\label{sec:con}
%%%%%%%%%%%%%%%%%%%%%%%%%%%%%%%%%%%%%%%%%%%%%%%%%%%%%%%

In this paper, we proposed a statistical learning-based one-bit ML detection method for uplink massive MIMO communications.
Since the performance of a learning-based one-bit detection approach can be severely degraded when the number of training samples is insufficient, the proposed method handled such challenges by adopting the adaptive dithering craft.
Without requiring channel knowledge, the dithering-and-learning method perform ML detection through learning likelihood functions at each antenna.
The proposed method is robust to the number of training symbols because the adaptive randomness triggers moderate fluctuation in the change of signs of the training sequence, thereby successfully extracting statistical pattern of one-bit quantized signals.
Simulation results demonstrate the detection performance of the proposed method in terms of SER.

\bibliographystyle{IEEEtran}
\bibliography{Learning_1bitML}

\begin{thebibliography}{10}
\providecommand{\url}[1]{#1}
\csname url@samestyle\endcsname
\providecommand{\newblock}{\relax}
\providecommand{\bibinfo}[2]{#2}
\providecommand{\BIBentrySTDinterwordspacing}{\spaceskip=0pt\relax}
\providecommand{\BIBentryALTinterwordstretchfactor}{4}
\providecommand{\BIBentryALTinterwordspacing}{\spaceskip=\fontdimen2\font plus
\BIBentryALTinterwordstretchfactor\fontdimen3\font minus
  \fontdimen4\font\relax}
\providecommand{\BIBforeignlanguage}[2]{{%
\expandafter\ifx\csname l@#1\endcsname\relax
\typeout{** WARNING: IEEEtran.bst: No hyphenation pattern has been}%
\typeout{** loaded for the language `#1'. Using the pattern for}%
\typeout{** the default language instead.}%
\else
\language=\csname l@#1\endcsname
\fi
#2}}
\providecommand{\BIBdecl}{\relax}
\BIBdecl

\bibitem{ngo2013energy}
H.~Q. Ngo, E.~G. Larsson, and T.~L. Marzetta, ``{Energy and spectral efficiency
  of very large multiuser MIMO systems},'' \emph{IEEE Trans. Commun.}, vol.~61,
  no.~4, pp. 1436--49, 2013.

\bibitem{larsson2014massive}
E.~G. Larsson, O.~Edfors, F.~Tufvesson, and T.~L. Marzetta, ``{Massive MIMO for
  next generation wireless systems},'' \emph{IEEE Commun. Mag.}, vol.~52,
  no.~2, pp. 186--95, 2014.

\bibitem{pi2011introduction}
Z.~Pi and F.~Khan, ``{An introduction to millimeter-wave mobile broadband
  systems},'' \emph{IEEE Commun. Mag.}, vol.~49, no.~6, 2011.

\bibitem{andrews2014will}
J.~G. Andrews, S.~Buzzi, W.~Choi, S.~V. Hanly, A.~Lozano, A.~C. Soong, and
  J.~C. Zhang, ``{What will 5G be?}'' \emph{IEEE Journal Sel. Areas in
  Commun.}, vol.~32, no.~6, pp. 1065--82, 2014.

\bibitem{heath2018foundations}
R.~W. Heath~Jr and A.~Lozano, \emph{Foundations of MIMO Communication}.\hskip
  1em plus 0.5em minus 0.4em\relax Cambridge University Press, 2018.

\bibitem{wen2016bayes}
C.-K. Wen, C.-J. Wang, S.~Jin, K.-K. Wong, and P.~Ting, ``{Bayes-optimal joint
  channel-and-data estimation for massive MIMO with low-precision ADCs},''
  \emph{IEEE Trans. Sig. Process.}, vol.~64, no.~10, pp. 2541--56, 2016.

\bibitem{studer2016quantized}
C.~Studer and G.~Durisi, ``{Quantized massive MU-MIMO-OFDM uplink},''
  \emph{IEEE Trans. Commun.}, vol.~64, no.~6, pp. 2387--99, 2016.

\bibitem{choi2017resolution}
J.~Choi, B.~L. Evans, and A.~Gatherer, ``{Resolution-adaptive hybrid MIMO
  architectures for millimeter wave communications},'' \emph{IEEE Trans. Sig.
  Process.}, vol.~65, no.~23, pp. 6201--16, 2017.

\bibitem{choi2018antenna}
J.~Choi, J.~Sung, B.~L. Evans, and A.~Gatherer, ``{Antenna Selection for
  Large-Scale MIMO Systems with Low-Resolution ADCs},'' \emph{IEEE Int. Conf.
  Acoustics, Speech, and Signal Process.}, 2018.

\bibitem{choi2021quantized}
J.~Choi, Y.~Cho, and B.~L. Evans, ``Quantized massive {MIMO} systems with
  multicell coordinated beamforming and power control,'' \emph{IEEE Trans.
  Commun.}, vol.~69, no.~2, pp. 946--61, 2021.

\bibitem{mezghani2007ultra}
A.~Mezghani and J.~A. Nossek, ``{On ultra-wideband MIMO systems with 1-bit
  quantized outputs: Performance analysis and input optimization},'' in
  \emph{IEEE Int. Symp. Info. Theory}.\hskip 1em plus 0.5em minus 0.4em\relax
  IEEE, 2007, pp. 1286--89.

\bibitem{mo2015capacity}
J.~Mo and R.~W. Heath, ``{Capacity analysis of one-bit quantized MIMO systems
  with transmitter channel state information},'' \emph{IEEE Trans. Sig.
  Process.}, vol.~63, no.~20, pp. 5498--5512, 2015.

\bibitem{wang2015multiuser}
S.~Wang, Y.~Li, and J.~Wang, ``{Multiuser detection in massive spatial
  modulation MIMO with low-resolution ADCs},'' \emph{IEEE Trans. Wireless
  Commun.}, vol.~14, no.~4, pp. 2156--68, 2015.

\bibitem{choi2015quantized}
J.~Choi, D.~J. Love, D.~R. Brown~III, and M.~Boutin, ``{Quantized Distributed
  Reception for MIMO Wireless Systems Using Spatial Multiplexing},'' \emph{IEEE
  Trans. Sig. Process.}, vol.~63, no.~13, pp. 3537--48, 2015.

\bibitem{choi2016near}
J.~Choi, J.~Mo, and R.~W. Heath, ``{Near maximum-likelihood detector and
  channel estimator for uplink multiuser massive MIMO systems with one-bit
  ADCs},'' \emph{IEEE Trans. Commun.}, vol.~64, no.~5, pp. 2005--18, 2016.

\bibitem{cho2019one}
Y.~Cho and S.-N. Hong, ``{One-Bit Successive-Cancellation Soft-Output (OSS)
  Detector for Uplink MU-MIMO Systems with One-Bit ADCs},'' \emph{IEEE Access},
  2019.

\bibitem{jeon2018one}
Y.-S. Jeon, N.~Lee, S.-N. Hong, and R.~W. Heath, ``{One-bit sphere decoding for
  uplink massive MIMO systems with one-bit ADCs},'' \emph{IEEE Trans. Wireless
  Commun.}, vol.~17, no.~7, pp. 4509--21, 2018.

\bibitem{park2021construction}
S.~Park, Y.~Cho, and S.~Hong, ``Construction of 1-bit transmit signal vectors
  for downlink mu-miso systems: Qam constellations,'' \emph{IEEE Trans. Veh.
  Techn.}, vol.~70, no.~10, pp. 10\,065--76, 2021.

\bibitem{mollen2017uplink}
C.~Moll{\'e}n, J.~Choi, E.~G. Larsson, and R.~W. Heath~Jr, ``{Uplink
  Performance of Wideband Massive MIMO With One-Bit ADCs},'' \emph{IEEE Trans.
  Wireless Commun.}, vol.~16, no.~1, pp. 87--100, 2017.

\bibitem{li2017channel}
Y.~Li, C.~Tao, G.~Seco-Granados, A.~Mezghani, A.~L. Swindlehurst, and L.~Liu,
  ``Channel estimation and performance analysis of one-bit massive {MIMO}
  systems,'' \emph{IEEE Trans. Sig. Process.}, vol.~65, no.~15, pp. 4075--89,
  2017.

\bibitem{nguyen2021svm}
L.~V. Nguyen, A.~L. Swindlehurst, and D.~H. Nguyen, ``{SVM}-based channel
  estimation and data detection for one-bit massive {MIMO} systems,''
  \emph{IEEE Trans. Sig. Proces.}, vol.~69, pp. 2086--99, 2021.

\bibitem{jeon2018supervised}
Y.-S. Jeon, S.-N. Hong, and N.~Lee, ``{Supervised-Learning-Aided Communication
  Framework for MIMO Systems with Low-Resolution ADCs},'' \emph{IEEE Trans.
  Veh. Technol.}, 2018.

\bibitem{jeon2018reinforcement}
Y.-S. Jeon, M.~So, and N.~Lee, ``{Reinforcement-learning-aided ML detector for
  uplink massive MIMO systems with low-precision ADCs},'' in \emph{IEEE
  Wireless Commun. and Networking Conf.}, 2018.

\bibitem{choi2019robust}
J.~Choi, Y.~Cho, B.~L. Evans, and A.~Gatherer, ``{Robust learning-based ML
  detection for massive MIMO systems with one-bit quantized signals},'' in
  \emph{IEEE Global Commun. Conf.}, 2019, pp. 1--6.

\end{thebibliography}

\end{document}